# Two-Dimensional Oscillatory Neural Network Based on Charge-Density-Wave Devices Operating at Room Temperature

A. Khitun, G. Liu, and A. A. Balandin

Department of Electrical and Computer Engineering, Bourns College of Engineering, University of California – Riverside, Riverside, CA 92521 USA

*Abstract* — We propose an oscillatory neural network implemented with two-dimensional tantalum disulfide devices operating in the change density wave regime at room temperature. An elementary cell of the network consists of two $1T-TaS_2$ devices connected in series. Such a cell has constant output and oscillatory states. All cells have the same bias voltage. There is constant current flowing through the cell in the constant output mode. The oscillations occur at a certain bias voltage due to the electrical-field driven metal-to-insulator transition owing to the changes in the charge density wave phase in the $1T-TaS_2$ channel. Two $1T-TaS_2$ devices oscillate out-of-phase where one of the devices is in the insulator phase while the other one is in the metallic state. The nearest-neighbor cells are coupled via graphene transistors. The cells are resistively coupled if the graphene transistor is in the On state while they are capacitively coupled if the transistor is in the Off state. The operation of the oscillatory neural network is simulated numerically for the 30×30 node network. The results of our numerical modeling show the formation of artificial vortexes and cellular-automata type data processing. The two-dimensional $1T-TaS_2$ devices, utilized in the network, offer a unique combination of properties such as scalability, high operational frequency, fast synchronization speed, and radiation hardness, which makes them promising for both consumer electronic and defense applications.

## I. INTRODUCTION

Presently, there is a strong motivation for development of a new generation of information processing systems, which function on the principles of biological or neuromorphic computing [1]. Such systems would drastically increase the computing efficiency in solving specific problems, particularly in image processing and pattern recognition. The neuromorphic computer, unlike the von Neumann computer, does not execute a list of commands, which constitute a program. Its major aim is not a general-purpose computation but rather special task data processing via the collective dynamics of the cells in the network [2]. The concept of a cellular neural network (CNN) was first formulated by Chua [3]. CNN is a two (three or more) dimensional array of mainly identical dynamical systems, called cells, which satisfy two properties: (i) most interactions are local within a finite radius, and (ii) all state variables are signals of continuous values. In subsequent works, the CNN paradigm evolved in many ways, and its computing capabilities in image processing and pattern recognition have been successfully demonstrated[4-7]. In recent years, the CNN concept attracted growing interest as a promising architecture for future information processing systems implemented with nanometer scale devices and structures [8]. Potentially, CNN comprising nano-cells will have a tremendous integration density, as well as specific architecture features originating from the unique output characteristics. A comprehensive review of nanoscale devices for the next generation computers is given in [9].

Oscillatory neural network is one of the promising approaches for development of the next generation computers. In such a network, an elementary cell comprises an oscillator circuit. The cells are locally coupled and may share a common node. The memorized and test patterns are encoded in the parameters of the oscillators. The collective behavior of the coupled oscillators is utilized for pattern recognition. There is a variety of nanoscale devices suitable for integration in an oscillatory network. An example of the oscillatory network with the coupled spin-torque oscillators is described in [10]. In general, an oscillatory network has two modes of operation: (i) the fixed point mode and (ii) the oscillatory mode. In the fixed point mode, the memories are stored as the fixed points of the network dynamics, whereas in the oscillatory mode they are encoded as the phase relations among individual oscillators.

In this work, we propose an oscillatory network based on two-dimensional (2D) tantalum disulfide (TaS$_2$) devices [11]. Specifically, we used 1T polytype of TaS$_2$, which undergoes the transition from a normal metallic phase to the charge density wave (CDW) phases at high temperature. The CDW state is a macroscopic quantum state formed via a periodic modulation of the electronic charge density accompanied by a periodic distortion of the atomic lattice in a quasi-1D or quasi-2D layered metallic crystal [12-15]. Some materials reveal several CDW phases with different transition temperatures. It is known that 1T-TaS$_2$ transforms from a normal metallic phase to an incommensurate CDW (IC-CDW) phase at 545 K, to a nearly commensurate CDW (NC-CDW) phase at 350 K and, finally, to a commensurate CDW (C-CDW) phase at 180 K [11]. Each phase transition is accompanied by a lattice reconstruction, which results in strong changes of the electrical properties of the material. The transition between the phases can be triggered by an applied voltage. We used 1T-TaS$_2$ as the device channel material where electrical current is switched by inducing transition between IC-CDW and NC-CDW states [11]. The integration of two 1T-TaS$_2$ devices provides a simple, miniaturized, voltage-controlled oscillator. A large number of 1T-TaS$_2$ oscillators can be integrated in a network, where the coupling between the oscillators can be controlled by graphene transistors.

The rest of the paper is organized as follows. In the next Section II, we outline the structure and fabrication of 1T-TaS$_2$ - graphene devices, and present experimental current-voltage (I-V) characteristics. In Section III, we describe the oscillatory neural network and present results of the numerical simulations illustrating the operation and collective behavior in the network. The experimental I-Vs are used as input data for our numerical model. The discussion and conclusions are given in sections IV and V, respectively.

## II. DEVICE STRUCTURE AND CHARACTERISTICS OF 1T-TaS$_2$ – BASED OSCILLATOR

Metal-insulator-transition (MIT) is the physical mechanism leading to the self-sustained oscillation. The simplest oscillatory circuit consist of one MIT device and a passive resistor as described in Ref. [16]. In order to demonstrate self-sustained oscillation in 1T-TaS$_2$ – based circuit, we build a prototype consisting of the 1T-TaS$_2$ channel connected in series with an off chip resistor, D-R configuration. The 1T-TaS$_2$ channel experiences switching between electrically conducting (IC-CDW) and resistive (NC-CDW) states as the voltage across the threshold value $V_{TH}$. The 1T-TaS$_2$ channel is capped with hexagonal boron nitride (h-BN) layer, which serves as the protective capping. The schematic of the oscillator device structure and an optical microscopy image of a typical fabricated device are shown in Figs. 1(A) and 1(B), respectively.

We fabricated a number of prototype devices for the proof-of-concept demonstration using mechanical exfoliation and transfer process. Briefly, the fabrication can be summarized in the following steps. Thin 1T-TaS$_2$ layer is exfoliated on SiO$_2$/Si substrate while BN thin layer is exfoliated from the bulk material onto the PDMS stamp. The stamp is then used to align and transfer BN on top of 1T-TaS$_2$ film [4-6]. In the proposed devices the metal contacts to 1T-TaS$_2$ are made in the edge contact configuration. The oscillations are observed when applying DC bias to the D-R circuit. As the voltage across 1T-TaS$_2$ channel exceeds $V_{TH}$, the oscillation occurs at the output port. The current-voltage characteristics of 1T-TaS$_2$ device and the oscillation waveform of the D-R circuit are shown in Fig.2. Load resistance is 1kΩ, DC bias is 4.21V, the frequency is 2MHz.

### III. Device Model and Results of the Numerical Modeling

We adopt the metal-insulator-transition (MIT) device model from Ref. [16]. The model has been validated with the experimental data. In the framework of this model, the device has two resistance states $R_H$ and $R_L$, where $R_H$ is the high resistance corresponding to the insulator state, and $R_L$ is the low resistance corresponding to the metallic state. The switching among the states is triggered by the voltage across the device. There are two threshold voltages $v_h$ and $v_l$ corresponding to the transitions between the metal and insulator phases. The resistance changes to a metallic ($R_L$) state as the voltage exceeds the higher threshold $v_h$. The change to the insulating ($R_H$) state occurs when the voltage exceeds the lower threshold $v_l$ . The thresholds $v_h$ and $v_l$ are not equal, i.e. there is hysteresis in the switching with $v_l < v_h$. Such a simple model describes well the I-V characteristics of a VO$_2$  MIT device [16].

The simplest way of building an oscillator circuit is to combine MIT device (D) with a resistor (R) in series as shown in Figure 3(A). The phase space of a single D-R oscillator is shown in Figure 3(B). As in Ref. [16], we define conductance $g_{di}=1/R_H$, $g_{dm}=1/R_L$, $g_S=1/R_S$, where subscript $d$ denotes a state dependent device conductance and $m/i$ denotes metallic/insulating state respectively, $R_s$ correspond to the linear resistor. Effective charging happens through $g_{dm}$ and whereas effective discharging happens only through $g_s$. The equation for the single D-R oscillator dynamics can be described by the following set of piecewise linear differential equations written as [16]:

$$cv' = \begin{cases} (v_{dd}-v)g_{dm} - vg_S & ch\arg ing \\ -vg_S & disch\arg ing \end{cases}$$

(1)

The D-R circuit produce oscillating output at certain combination of resistances $R_H, R_L$, $R_S$ and bias voltage $v_{dd}$, as explained in Figure 1(B). Lines with slopes $R_H$ and $R_L$ are the regions of operation of the device in insulating and metallic states respectively. The intersection of these lines with the load line due to the series resistance $R_S$ gives the stable points of the system in the two states. In order to obtain self-sustained oscillations, the stable points in each state should lie outside the region of operation, i.e. outside the region defined by vertical lines passing through the transition points. In this case, MIT device has phase transition before the circuit reaches the stable point. Figure 3(C) shows the results of numerical modeling illustrating the oscillating output of the D-R oscillator. Hereafter, we depict all values in normalized units $R_0, I_0,$ and $V_0$ where $I_0=V_0/R_0$. In our numerical simulations, we use the following set of parameters $v_l = 1 \ V_0$, $v_h = 2 \ V_0$, $R_H = 2.73 \ R_0$, $R_L = 0.67 \ R_0$, $R_S = 1 \ R_0$.

Following the same methodology, one can write the equation for D-D oscillator circuit as shown in Fig 4(A):

$$cv' = \begin{cases} (v_{dd}-v)g_{1dm} & ch\arg ing \\ -vg_{2dM} & disch\arg ing \end{cases}$$

(2)

The only difference from the D-R case is that the effective charging happens through $g_{1dm}$ and effective discharging through $g_{2dm}$. The conditions leading to the self-sustained oscillation are illustrated in Fig. 3(B). Due to the symmetry in *I-V* characteristics, two MIT devices are biased in such a way, that the phase transition in each device occurs before the system reaches one of the four stable points. In this scenario, as one of the MIT devices changes its state from metallic to insulator, the second MIT device has phase transfer from insulator to metal state, and vice versa. As a result, the system of two MIT devices forms a *complementary pair*, where one of the devices is in high resistance state, while the other is in low resistance state. Fig. 4(C) shows the results of numerical modeling illustrating the switch between the resistance states for MIT devices in D-D configuration. Compare to the R-D configuration, D-D circuit possesses a lower leakage current similar to the well-established complementary metal oxide semiconductor (CMOS).

There are different possible ways to oscillator coupling. There are two extreme cases of purely resistive and purely capacitive coupling. In the case of purely resistive coupling, two oscillators are tending to oscillate in phase. In contrast, purely capacitive coupling makes the two oscillators to oscillate out of phase. The dynamics of purely resistive and purely capacitive coupled D-D oscillators are described in details in Ref. [16]. In this work, we propose the combination of MIT-based oscillators and a graphene transistor as shown in Fig. 5. Two D-D circuits are coupled via resistance $R_C$ and capacitance $C_C$. The mutual capacitance $C_c$ is fixed after the fabrication, while the coupling resistance $R_C$ is controlled by the state of the graphene transistor (G-FET). In the On state, the resistive coupling dominates forcing the two oscillators to oscillate in phase. In the Off state, the capacitive coupling comes to the stage by introducing a π-phase shift between the oscillators. It is important to note, that two capacitively coupled D-D oscillators may occur in a meta-stable state (i.e. in-phase oscillation) till an external perturbation will make the system to evolve towards the stable out-of-phase state. The introduction of a transistor controlling the coupling between the MIT-based oscillators in a network opens a number of intrigue possibilities for engineering special templates for data processing. In the rest of this work, we show some examples of the proposed approach

As a testbed, we consider a 30×30 template of identical D-D coupled oscillators (i.e. as shown in Fig. 4). An elementary cell of the network is shown in Fig. 6, where a D-D oscillator is connected to the four nearest-neighbor D-D oscillators. The position of the cell in the network is defined by the two numbers Nx and Ny. The state of the cell is assigned to the voltage $V(N_X,N_Y)$ as depicted in Fig.6(B). There are four junction transistors connecting the central oscillators with the four neighbors. As in Fig.5, the states of these transistors define the type of coupling (e.g. the On state of the transistor corresponds to the inter-cell resistance of $0.1R_0$, the Off state corresponds to the inter-cell resistance $10 R_0$). For instance, oscillator $(N_X,N_Y)$ can be capacitively coupled to the oscillator $(N_{X+1},N_Y)$ but resistively coupled to the other neighbors.

Let us consider a template with two regions consisting of capacitively and resistively coupled oscillators. The capacitively coupled oscillators are located in the center of the template (i.e. all oscillators around (15,15) with radius 7). All the oscillators outside this region are resistively coupled. We assume all the oscillators to be biased with the same voltage $V_{DD}=3V_0$. Such a network demonstrates an interesting dynamics, which is illustrated by numerical modeling.

The color surface in Fig. 7 presents the results of numerical modeling. The color surface shows the voltage map in the 2-D array of oscillators. Nx and Ny depict the position of the oscillator in the network. The color depicts the voltage of the cell. The resistively coupled oscillator oscillate in phase (i.e. the same voltage in every moment of time), while the capacitively-coupled oscillators in the center oscillate out-of-phase. In Fig. 8, we show the sequence of voltage maps calculated with the time interval of $0.05t_0$, where $t_0=R_0C_0$, where $C_0 =0.01$. This dynamics resembles a magnetic vortex, where the position of magnetization is copied by the phase of oscillation. In contrast to the steady-state magnetic vortexes, this artificial vortex in the sea of coupled oscillators does not show any stead-state phase distribution. The phase difference between any oscillator in the center and any oscillator out of the central region varies with time. This phase evolution is defined by the set of electric characteristics as well as the geometry of the capacitively-coupled region.

In Fig. 9, we present the results of numerical simulations showing signal propagation through the network with partially capacitive and resistive-coupled oscillators. The initial distribution of the cell voltages is shown in top left graph (i.e., t=0) . All of the cells except one (2,15) are prepared in the meta-stable state oscillating in-phase regardless of the coupling. The oscillator at (2,15) provides the input, which triggers the switch from the metastable in phase to stable out-of-phase coupling for all capacitive coupled oscillators. This switching is very well seen in the sequence of snapshots taken with time interval depicted in $t_0$. The propagation of the signal is associated with the sequential change of the phase of oscillation for capacitive-coupled oscillators. For instance, it shows the switch of the oscillation phase in the center circular region around (Nx=15,Ny=15), with capacitevely coupled cells. As the signal propagates through the template, one can see the phase ornament produced by the capacitively-coupled oscillators.

It is interesting to note the similarity between the operation of the array of capacitively-coupled oscillators and magnetic cellular automata [17]. In both cases, the bulk of the elementary cells are prepared in a metastable state. Signal propagation is resulted in the cell relaxation. In magnetic cellular automata, the nearest-neighbor magnets tend to be polarized in the opposite directions due to the dipole-dipole interaction. In the network built of oscillators, the nearest-neighbor capacitively coupled oscillator trend to oscillate out-of-phase. Potentially, the array of coupled oscillators can perform the same basic logic operations such as MAJ and NOT gate as magnetic cellular automata[18].

Finally, we present the results of numerical modeling showing the Game of Life in the network of coupled 1T-TaS$_2$ oscillators The Game of Life is a special version of cellular automation invented by the J.H. Conway in 1970 [19]. It consists of a collection of cells which may have two or more states (e.g. black and white). The cells change their states according to the set of rules, which results in a system evolution mimicking living cell birth, multiplication and death. Depending on the initial conditions, the cells may form various patterns throughout the evolution. Game of Life offers an original way of implementing complex Boolean operators in the space-time dynamics and presents a testbed for artificial intelligence development.

## IV. DISCUSSION

There are several technological advantages inherent to 1T-TaS$_2$ – graphene devices including scalability, high operation frequency, fast synchronization, voltage control of operational frequency, flexibility, radiation hardness and room temperature operation. It is well known that G-FETs show relatively low $I_{on}/I_{Off}$ ratio due to the absence of a band gap, which is significantly narrows the perspectives of using G-FETs in conventional digital logic [20] . In our proposed architecture, G-FETs are utilized for the oscillatory circuits coupling, where the $I_{on}/I_{Off}$ ratio of 100 is sufficient for switching between the resistive and capacitive regimes.   At the same time, G-FETs can operate at very high frequencies of hundreds of GHz [21, 22]. The transitions between IC-CDW and NC-CDW phases in 1T-TaS$_2$ channel can reach frequencies of ~1 THz [23]. These considerations suggest that the frequency of operation of 1T-TaS$_2$ – G-FET devices can be increased substantially compared to the one demonstrated experimentally. In the present experiments, the frequency is limited by the extrinsic RC time constants of the probe station measurement apparatus. The intrinsic resistances and capacitances can be reduced by scaling. It is important to note that 1T-TaS$_2$ – based oscillator circuits do not require any micro- or nano-inductor components, which is one of the major issue for integrated magnetics [24].  Additional advantage of the selected material systems is that graphene and h-BN layers in the device structure have extremely high thermal conductivity[25, 26], and, as a result, act as heat spreaders. Graphene is a very promising candidate for applications in flexible electronics due to its high carrier mobility and mechanical flexibility [27]. The combination of graphene transistors with 2D 1T-TaS$_2$ may pave a road towards the new generation of flexible circuits. For some applications, e.g. deep-space probes, radiation hardness is an important metric. The considered 1T-TaS$_2$ – G-FETs are all-metallic devices, which are less susceptible to radiation damage than semiconductors. There have been a number of attempts to implement all-metallic switches and circuits, ranging from metal dot single electron transistors, to metallic carbon nanotube devices, and metallic nanowire transistors[28-32]. Our CDW devices belong to the same class of devices that do not use semiconductor components. The results of our numerical modeling show fast synchronization among the coupled oscillators. It takes only a couple of cycles for the 1T-TaS$_2$ – graphene device for transition from the in-phase to the out-of-phase oscillatory states. This is another important benefit of the proposed technology.

 The presented data suggests that overall, the unique output characteristics of 1T-

TaS$_2$ devices are of great potential for implementation in phase-based logic devices using oscillatory nano-systems [33]. Phase logic was invented by Eiichi Goto and John von Neumann in the middle of the last century [34, 35]. The key idea of this approach is to exploit two (or more) distinct oscillating steady states for logical states representation. The first phase-based logic devices were built more than sixty years ago primarily in Japan and competed for some time with digital logic circuits. Later on, phase-based logic devices left the stage due to the great success of transistors and integrated circuits. Scalability and low operational speed are the two main disadvantages inherent to von Neumann's schemes. The need in large inductors and capacitors limits the size of the oscillatory circuits. Also, periodic turn-on transients and delays in logic gates make phase-based logic slower than transistorized level-based logic. The development of nanometer systems together with the implementation of novel physical mechanisms (e.g. spin torque oscillator [36]) has led to the growing interest in phase-logic devices able to overcome the limits of von Neumann's schemes.

## V. CONCLUSIONS

We presented a design of an oscillatory neural network based on integrated 1T-TaS$_2$ – graphene devices. The oscillations are driven by 1T-TaS$_2$ channel switching between two CDW states. Graphene FETs are used for coupling between the 1T-TaS$_2$-based oscillatory cells. . We illustrated the dynamics of the proposed network by numerical modeling of a network comprising 32×32 cells, where each cell consists of two 1T-TaS$_2$ devices. The type of coupling (e.g. capacitive or resistive) between the nearest-neighbor cells is controlled by the G-FETs. The results of numerical simulations reveal potential of the proposed network for general and special task data processing. The integrated 1T-TaS$_2$ – graphene devices possess a unique combination of properties such as scalability, high operational frequency, fast synchronization speed, radiation hardness, which makes well suited for application in oscillatory networks.

**ACKNOWLEDGMENT**

This work was supported, in part, by the National Science Foundation (NSF) Emerging Frontiers of Research Initiative (EFRI) 2-DARE project Novel Switching Phenomena in Atomic $MX_2$ Heterostructures for Multifunctional Applications (NSF 005400). The 1T-$TaSe_2$ crystals used for fabrication of the prototype devices were provided by Professor Tina T. Salguero, University of Georgia.

**Figure Captions**

Figure 1. (A) Structure of 1T-TaS2 device. 1T-TaS2 thin film is fully covered with h-BN, which acts as a protective layer against oxidation. (B) The optical image of a typical 1T-TaS2 device.

Figure 2. (A) Current-voltage characteristics of 1T-TaS2 device at room temperature. The CDW phase transition can be triggered by applying voltage beyond the threshold level. (B) Oscillation waveform of a D-R circuit. The load resistance is 1kΩ, DC bias is 4.21V.

Figure 3. (A) Schematics of an oscillator circuit realized with a MIT device in series with a resistor (D-R configuration). (B) Phase space of the device in a single (D-R) oscillator. (C) Results of numerical modeling showing the output of the (D-R) oscillator.

Figure 4. (A) Schematics of relaxation oscillator circuit realized with two MIT devices connected in series - (D-D) configuration. (B) Phase space of the D-D oscillator. (C) Results of numerical modeling showing the output of the (D-D) oscillator. (D) Resistance states of the 1T-TaS2 devices as a function of time. The devices are mostly in the different resistance states (i.e. if one of the devices is in the high-resistance state, the second one is in the low resistance state).

Figure 5. Schematics of two D-D coupled oscillators. The type of coupling is controlled by the transistor. In the On state, the resistive coupling dominates forcing the two oscillators to oscillate in phase. In the Off state, the capacitive coupling comes to the stage by introducing a π-phase shift between the oscillators.

Figure 6. (A) Schematics of 2D network consisting of D-D coupled oscillators. (B) Elementary cell of the network comprising nearest-neighbor coupled oscillators. The coupling between the oscillators is controlled by the junction transistors

Figure 7. Results of numerical simulations. The color surface shows the voltage map in the 2-D array of oscillators. Nx and Ny depict the position of the oscillator in the network. The oscillators in the center of the network around Nx=15, Ny=15 with radius 7, are capacitively coupled. The rest of the oscillators are resistively coupled.

Figure 8. Results of numerical simulations showing the dynamics in the template with capacitively- (center) and resistively- (out of the central region) coupled oscillators. The graphs (A-E) are the snapshots of the network voltage map taken with the time step 0.05[RC].

Figure 9. Results of numerical modeling illustrating Cellular Automata-like network operation. (A) The initial distribution of the cell voltages. All of the cells except one (2,15) are prepared in the meta-stable state oscillating in-phase regardless of the coupling. (B) and (C) signal propagation is associated with the switch of the oscillation phase in the center circular region around (Nx=15,Ny=15).

Figure 10. Results of numerical simulations showing system evolution. (A) Starting configurations. There is a group of capacitively-coupled oscillators in the left bottom corner – cell alive. All other oscillators oscillate in-phase – dead cells. (B) System evolves by driving more capacitively-coupled oscillators out of the metastable state. (C) Culmination of living cell growth. As the living cells reach the top right corner, it triggers the coupling switch for the oscillators in the left bottom corner. The oscillators become resistively coupled. (D) The oscillators in the left bottom corner oscillate in-phase, which symbolizes cell death.

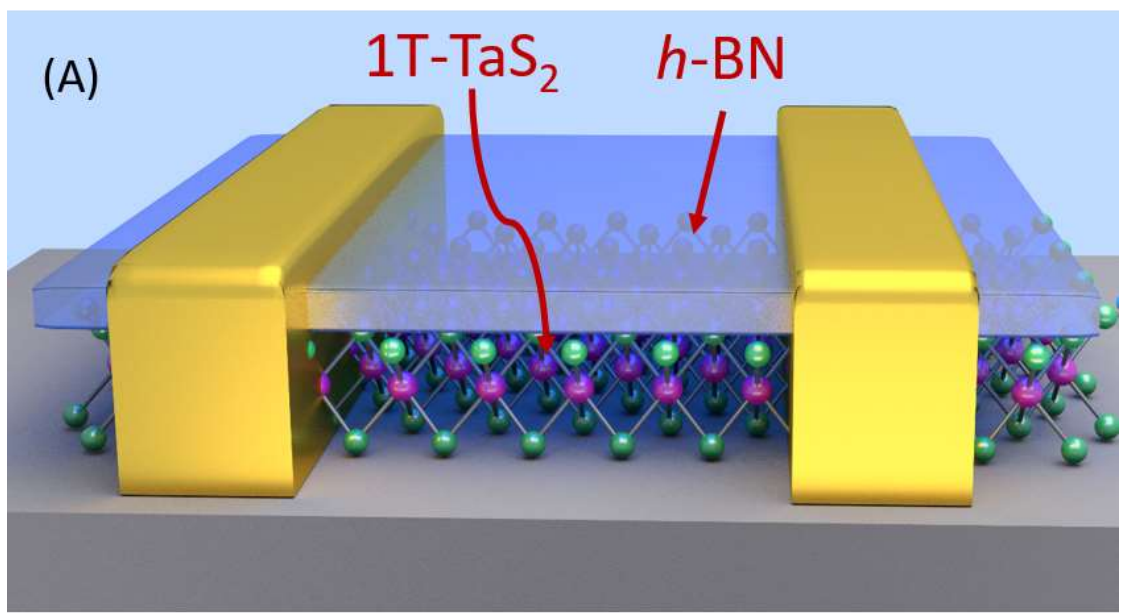

(A) 1T-TaS$_2$    *h*-BN

(B)

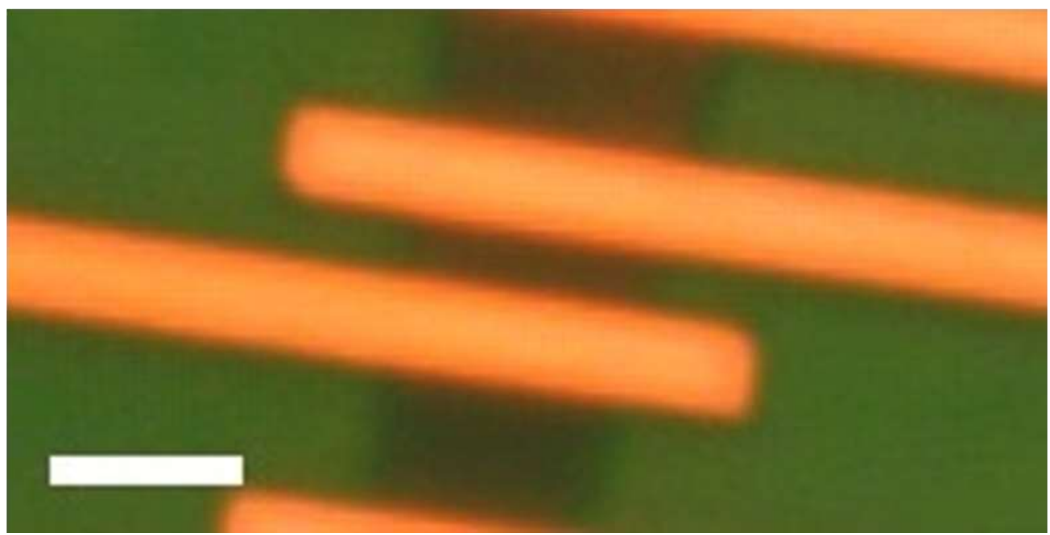

Figure 1

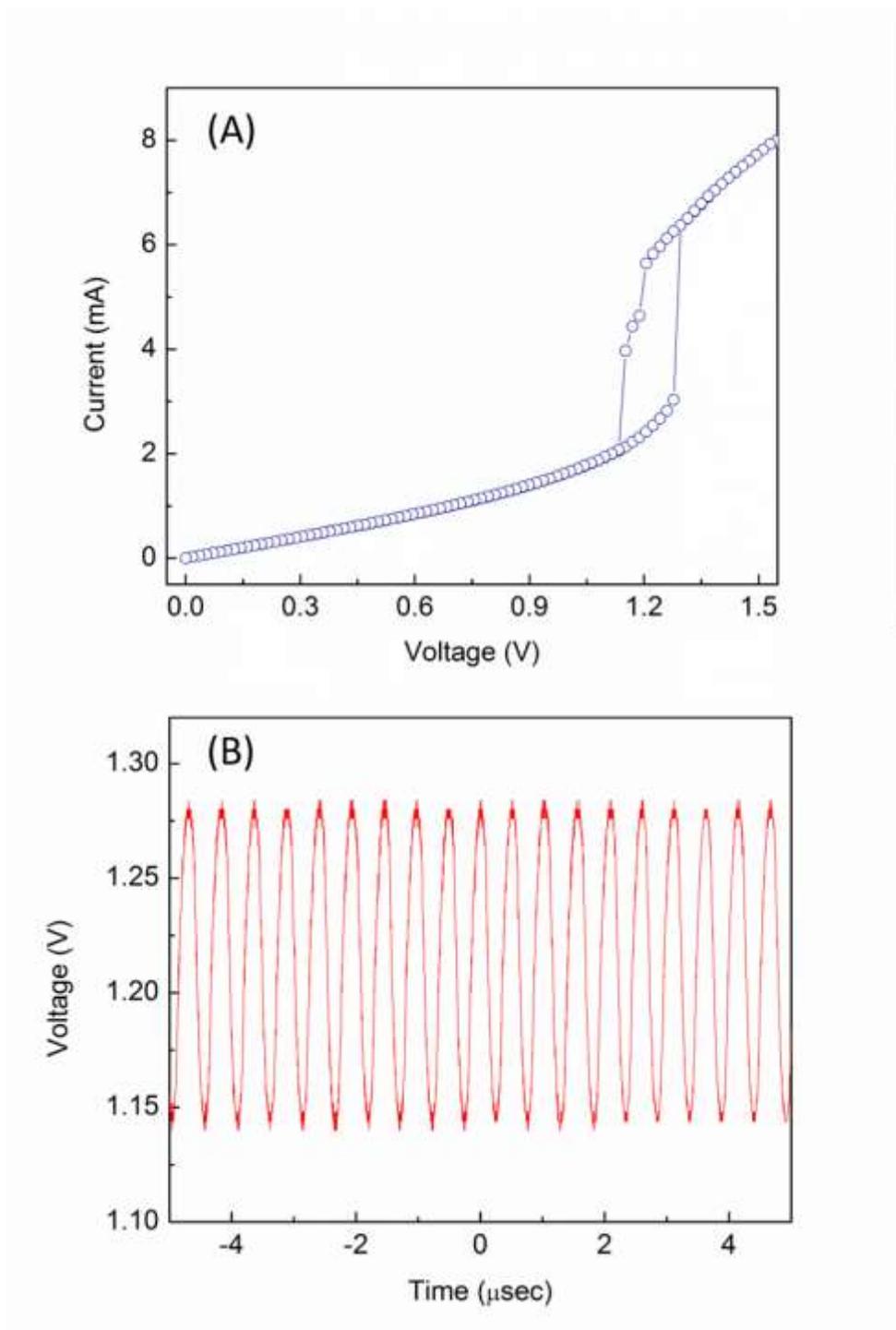

Figure 2

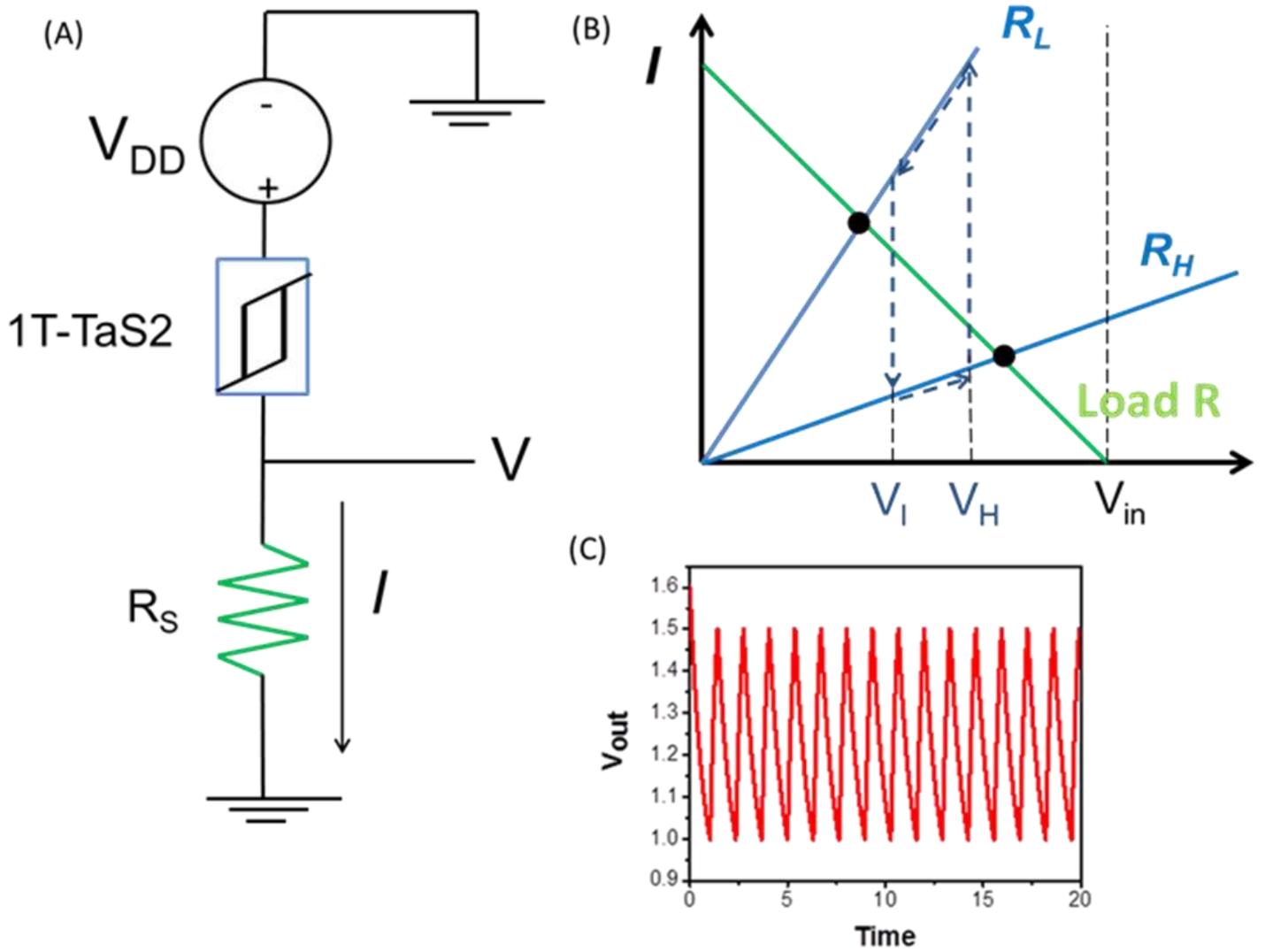

Figure 3

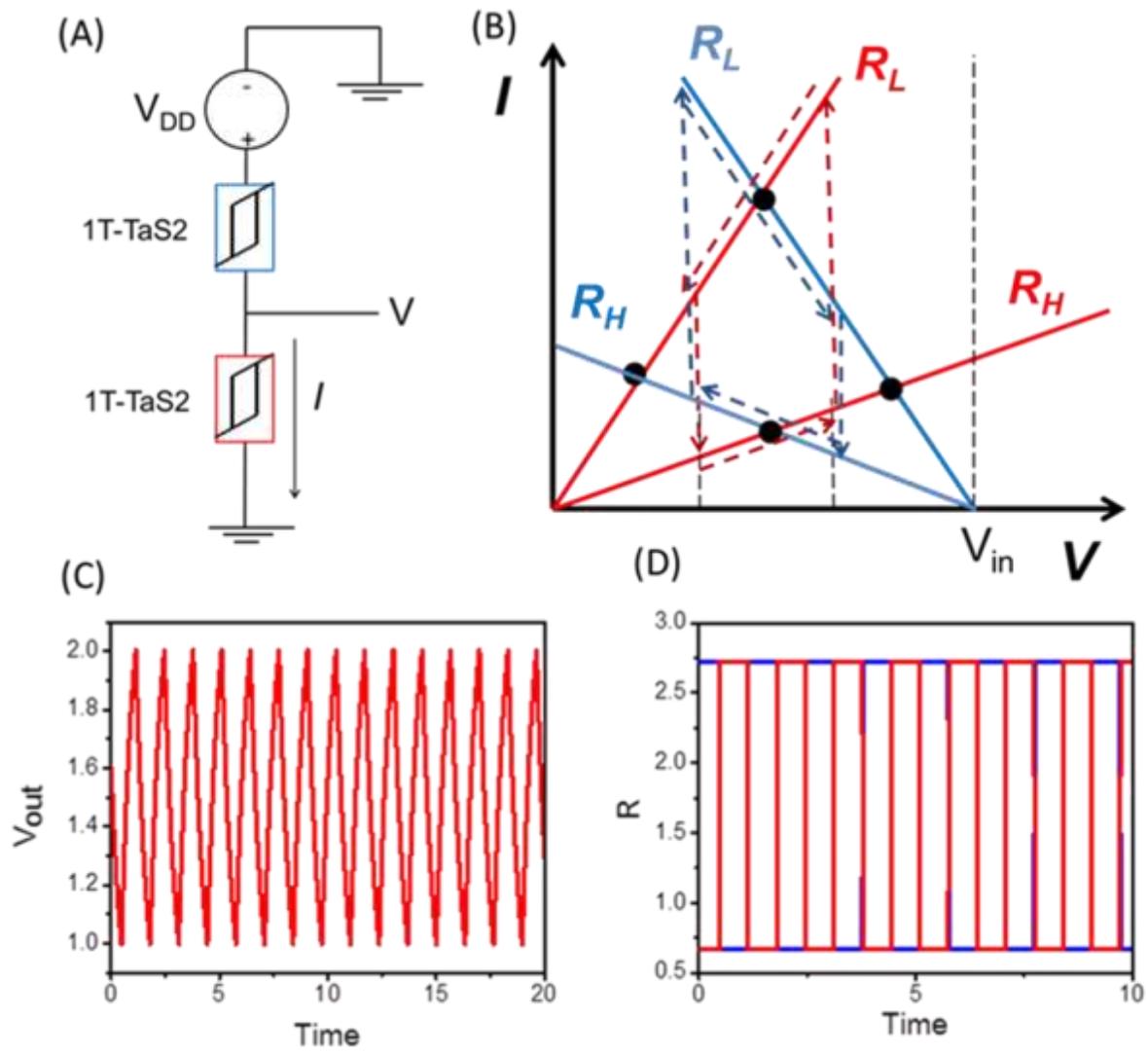

Figure 4

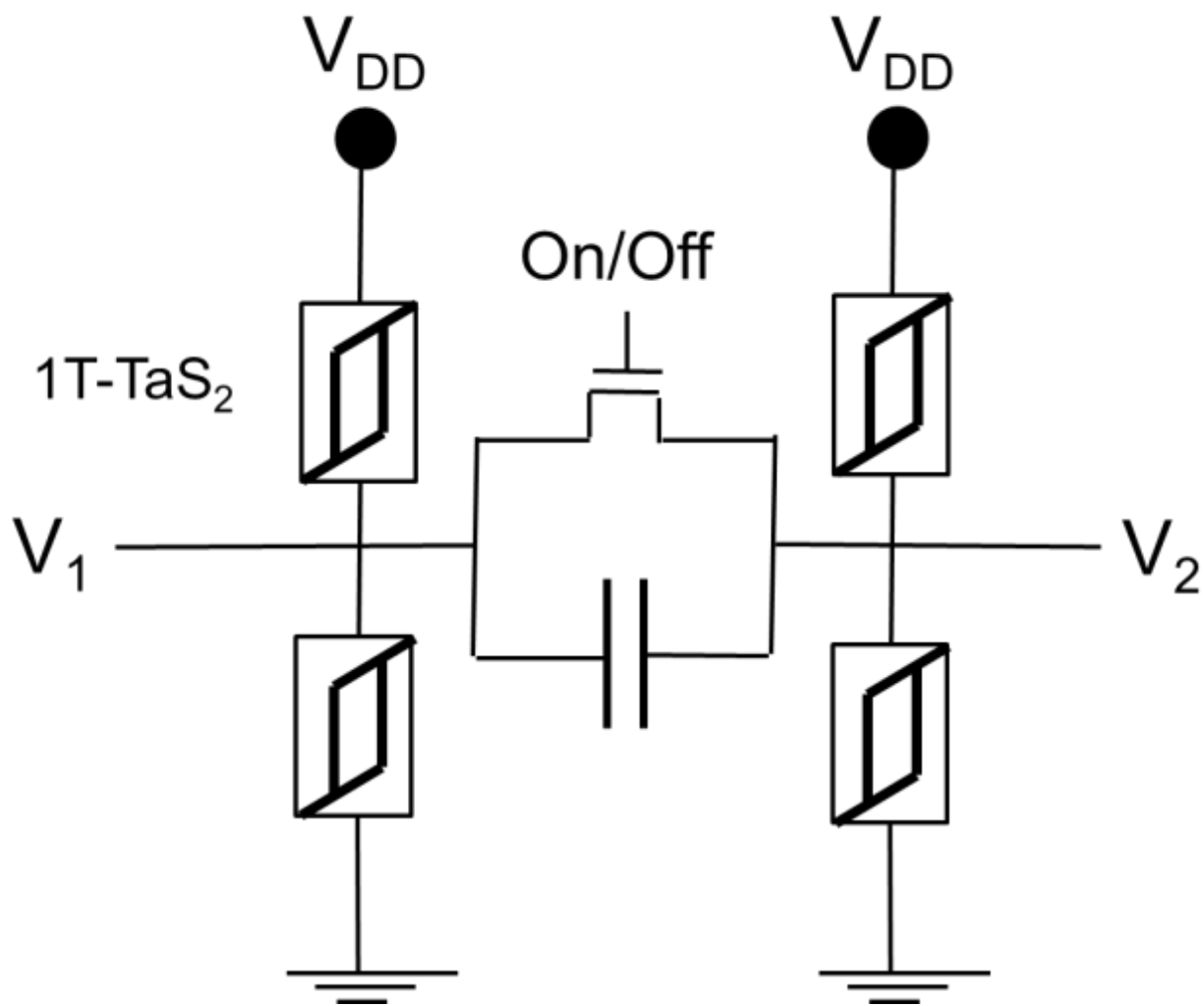

Figure 5

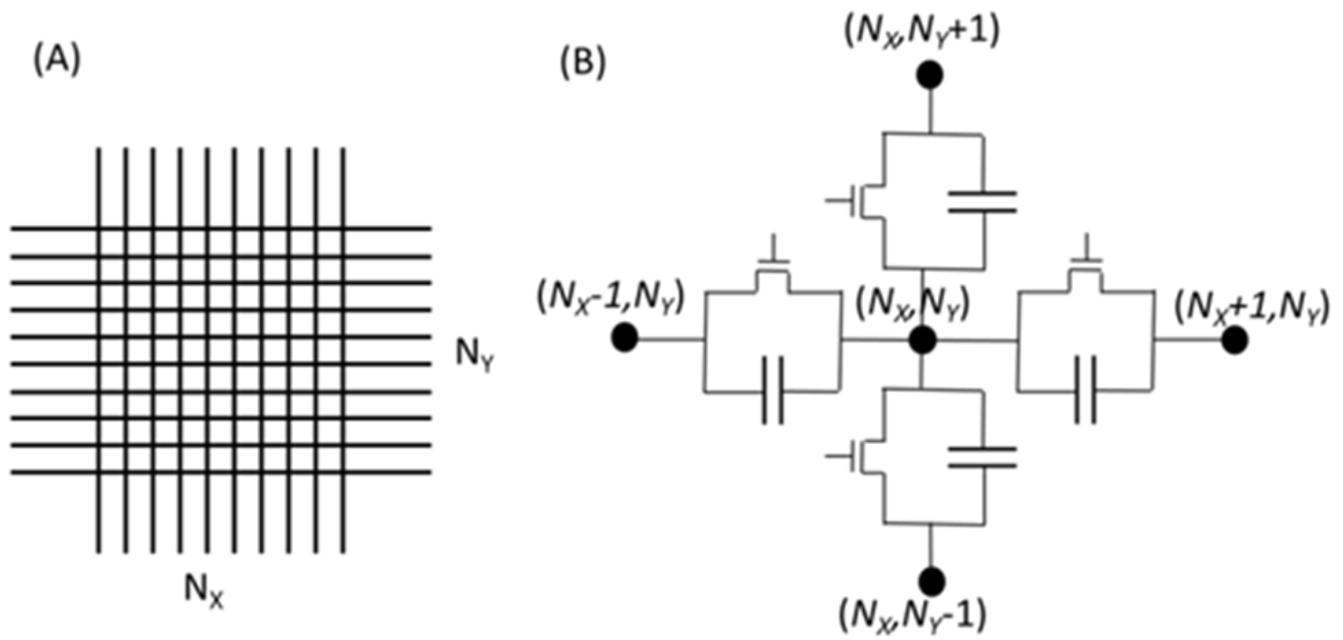

Figure 6

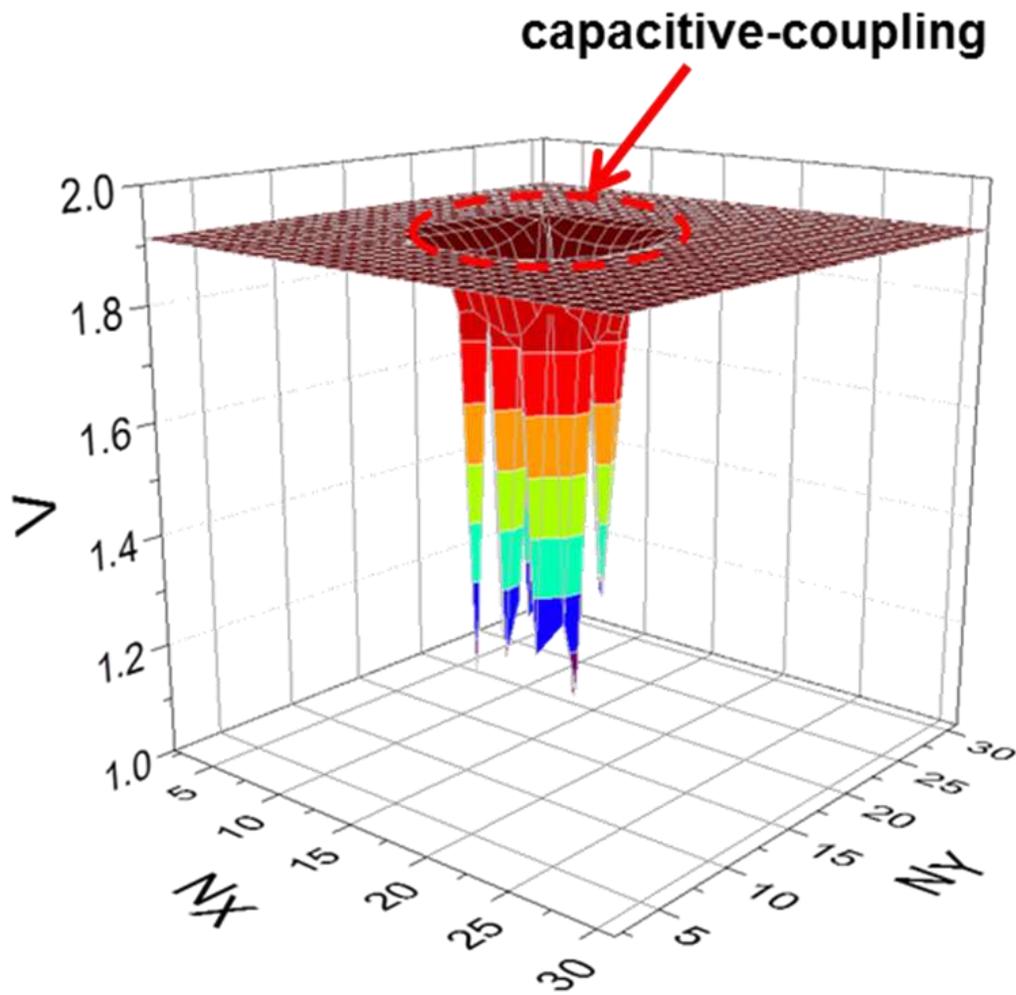

Figure 7

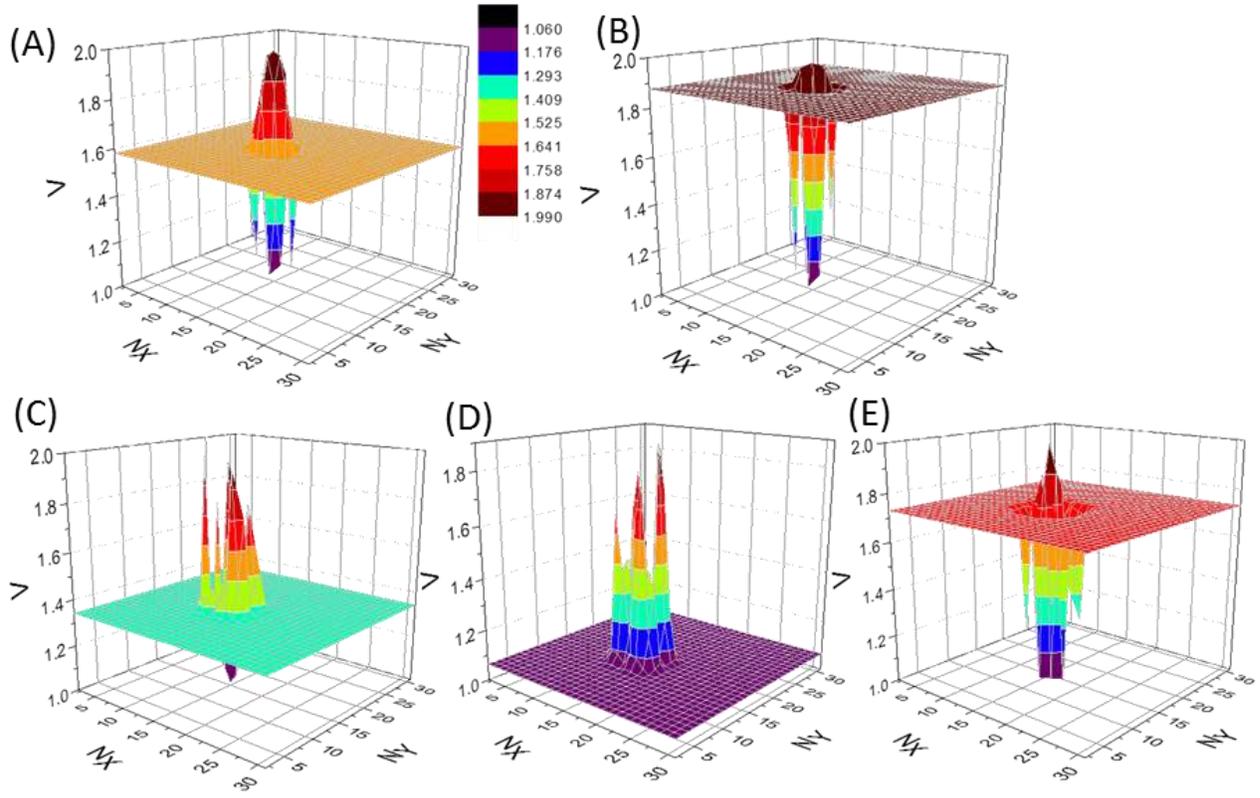

Figure 8

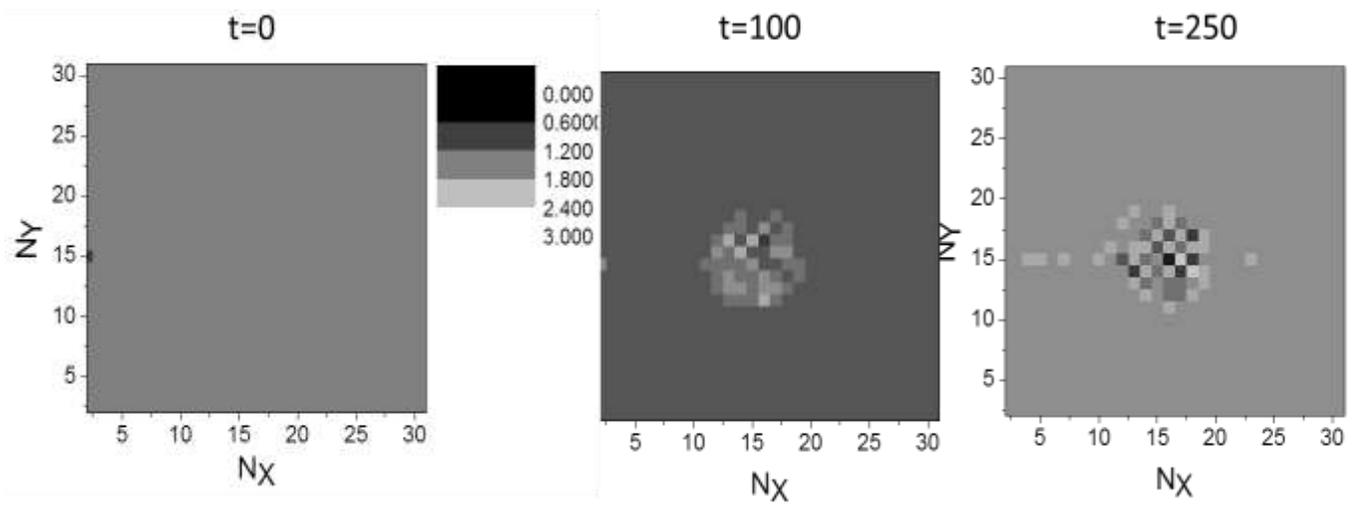

Figure 9

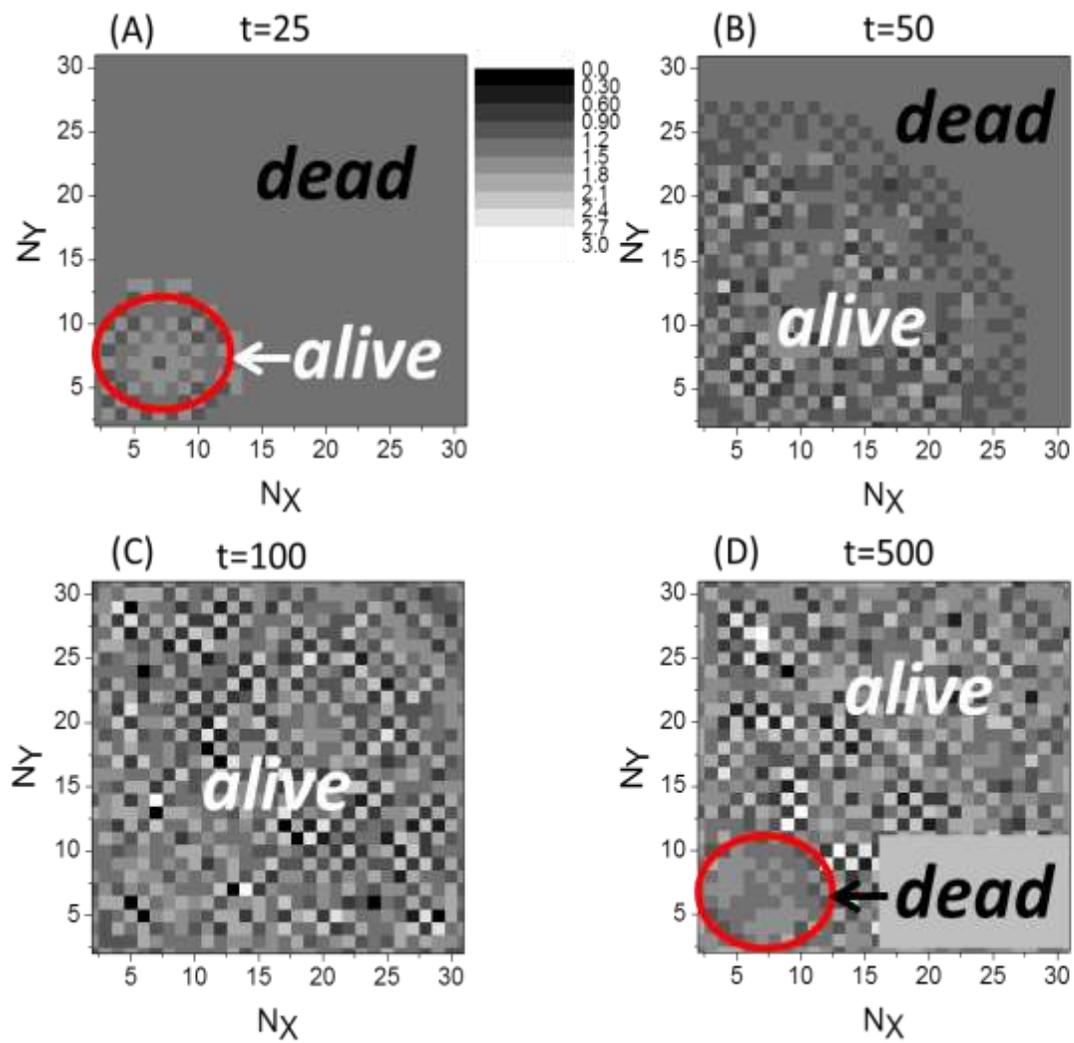

Figure 10